\documentclass[a4paper]{jpconf}
\usepackage{graphicx}
\begin{document}

\def\lapproxeq{\lower .7ex\hbox{$\;\stackrel{\textstyle                                                    
<}{\sim}\;$}}                                                    
\def\gapproxeq{\lower .7ex\hbox{$\;\stackrel{\textstyle                                                    
>}{\sim}\;$}}                                                    
\def\be{\begin{equation}}                                                    
\def\ee{\end{equation}}                                                    
\def\bea{\begin{eqnarray}}                                                    
\def\eea{\end{eqnarray}}
\def\b{\vec{b}}
\def\bb{\vec{b}'}     
\def\q{\vec{q}} 
\def\pp{\vec{p}'_t}                                                    
\def\k{\vec{k}_t} 
\def\kk{\vec{k}'_t} 
\def\GeV{\rm GeV}
\begin{flushright}                                                    
IPPP/14/19  \\
DCPT/14/38 \textbf{}\\                                                    
\today \\                                                    
\end{flushright} 

\vspace{-2.0cm}
\title{Lessons from LHC elastic and diffractive data\footnote{To be published in the Proceedings of the International Workshop on Particle Physics Phenomenology in memory of Alexei Kaidalov, Moscow, 21-25 July, 2013.}}

\author{Alan D. Martin}

\address{Institute for Particle Physics Phenomenology, University of Durham, Durham, DH1 3LE }

\ead{a.d.martin@durham.ac.uk}

\begin{abstract}

We discuss a model which gives a `global' description of the wide variety of high-energy elastic and diffractive data that are presently available, particularly from the LHC experiments. The model is based on only one pomeron pole, but includes multi-pomeron interactions. Significantly, the LHC measurements require that the model includes the transverse momentum dependence of the intermediate partons as a function of their rapidity, which results in a rapidity (or energy) dependence of the multi-pomeron vertices.

\end{abstract}

\section{Introduction}

Since this Workshop is in memory of Aliosha Kaidalov, it is appropriate to list some of his pioneering contributions to the understanding of Diffractive processes in high energy hadron interactions. He was the first to evaluate the effects of low-mass diffractive dissociation, namely processes of the form $pp \to p+N^*$ \cite{Kaidalov7}. Next, Kaidalov et al. \cite{KKPT} were the first to perform a triple-Regge analysis, which underlies the description high-mass dissociation. Together with Ter-Martirosyan, Aliosha proposed \cite{Kaidalov6} that multiple hadron production at high energies was the result of the creation and breaking of quark-gluon strings in hadron-hadron collisions. It is shown that this approach, together with the general results of supercritical Pomeron theory ($\alpha_P(0)>1$), yields a natural description of all high-energy multiple production data. Again with collaborators, Kaidalov et al. \cite{Kaidalov3,Kaidalov4} pioneered a model for soft high energy interactions, which included multi-pomeron diagrams in a global description of the data, for the first time. These contributions underlie the models developed for understanding diffractive processes to this day: here we may mention the Durham \cite{KMR-s2,KMR-s3} and Tel-Aviv \cite{Maor,Gotsman} models, the work by Ostapchenko \cite{Ost} and by Poghosyan and Kaidalov \cite{KaidalovPog}.

Basically the models describe high-energy diffractive processes  by pomeron exchange
within the framework of Reggeon Field Theory (RFT)~\cite{Gribov}.  Elastic and low-mass diffractive dissociation are described in terms of a multi-channel eikonal and high-mass diffraction in terms of diagrams with multi-pomeron couplings, the simplest of which is the triple-pomeron diagram.  

Recently a wide variety of measurements of diffractive processes has been obtained by experiments at the LHC. Attempts at a `global' simultaneous description of these data show differences with the expectations of conventional Regge Theory. Here, we explain how, in Refs. \cite{kmr2,rev}, it has been possible to modify the classic RFT in a physically-motivated way so as to accommodate the tendencies observed at the LHC. The crucial modification is to include the transverse momentum dependence of the intermediate partons on rapidity (or energy).  We may call this the $k_t(s)$ effect.  But, first, we list the high-energy diffractive data that exist at present.  Then we outline some of the problems presented by the data.

\section{The high-energy diffractive data   \label{sec:data}}
Quite a wide variety of diffractive measurements have been made at the LHC.
At the moment, data for diffractive processes are available at 7 TeV. The most detailed data come from the TOTEM collaboration. TOTEM have measured\\ (i) the total and elastic cross sections (in a wide $t$ interval including the dip region)~\cite{TO1,TO}; \\(ii) the cross section of low-mass ($M_X <3.4$ GeV) 
 proton dissociation \cite{TO2}; \\(iii) double dissociation ($pp\to X_1+X_2$)~\cite{TO3}; and \\(iv) made preliminary measurements of high-mass single proton dissociation, $\sigma_{\rm SD}$, integrated over three intervals of $M_X$: namely $(3.4,8);~(8,350);~(350,1100)$ GeV~\cite{TO4}. \\In addition, we have the inelastic cross sections and the cross sections of events with a Large Rapidity Gap (LRG) measured by the ATLAS \cite{atl}, CMS \cite{CMSdiff} and ALICE \cite{ALICE} collaborations.  Moreover, we have valuable data on elastic and proton dissociation from experiments at the Tevatron \cite{EEE,cdfB,GM}.

\section{Potential puzzles in the diffractive data sets   \label{sec:items}}
Following \cite{kmr2,rev}, we list several potential puzzles that seem to exist in the data. 
\begin{itemize}
\item[(a)] First, we note that the total cross section in the Tevatron -- LHC energy interval starts to grow {\it faster}, not more slowly, than below the Tevatron energy, and, secondly, the slope of the effective pomeron trajectory, $\alpha'_{\rm eff}$, increases \cite{Schegelsky}. 
In particular, 
Donnachie-Landshoff-type fit \cite{DL} predicts $\sigma_{\rm tot}=90.7$ mb at $\sqrt{s}=7$ TeV, while TOTEM
observes 98.6$\pm 2.2$ mb \cite{TO}. Moreover, the elastic slope was measured
at the Tevatron ($\sqrt s=1.8$ TeV) to be $B_{\rm el}=16.3\pm 0.3$ GeV$^{-2}$ by
the E710  experiment \cite{EEE} and to be
$B_{\rm el}=16.98\pm 0.25$ GeV$^{-2}$ by the CDF group \cite{cdfB}. Even starting from
the CDF result, and using the $\alpha'_P=0.25$ GeV$^{-2}$, we expect
$B_{\rm el}=16.98+4\times 0.25\times \ln(7/1.8)=18.34~ \GeV^{-2}$  at 7 TeV, while TOTEM finds $19.9\pm 0.3$ GeV$^{-2}$ \cite{TO}.
  Also, in the relatively low $|t|<0.3\ -\ 0.4$ GeV$^2$ region
we do not see the expected increase of $t$-slope as $|t|$ increases.%
\item [(b)] The experimental information on low-mass dissociation  is a puzzle in that the cross section
$\sigma_{\rm D}^{{\rm low}M_X}$ goes from about $2-3$ mb at the CERN-ISR energy
of 62.5 GeV to only $2.6\pm 2.2$ mb at 7 TeV at the LHC \cite{TO2}. Thus $\sigma_{\rm D}^{{\rm low}M_X}$ is about 30$\%$ of $\sigma_{\rm el}$ at 62.5 GeV and only about 10$\%$ at 7 TeV, whereas we would expect these percentages to be about the same for single pomeron exchange.
\item [(c)] An analogous problem occurs for high-mass dissociation where the cross section $d\sigma_{\rm SD}/d\ln\xi$ in the first (3.4 to 8 GeV) $M_X$ interval is more than twice larger than that in the central interval. Of course, according to the triple-Regge formula, a pomeron intercept $\alpha_P(0)>1$ leads to an increase of the cross section when $\xi$ decreases, but by the same argument we have to observe a larger cross section at the LHC than at the Tevatron, for the same value of $M_X$, contrary to the data. 
\item [(d)] The `factorisation' relation between the observed elastic, single and double proton dissociation cross sections is intriguing, and appears not easy to explain.
\end{itemize}
In Refs. \cite{kmr2,rev} it is claimed that a global description, including the $k_t(s)$ effect, is able to account for all these puzzles in the data. This will be the subject of Section \ref{sec:kts}.

\section{Eikonal approach to elastic scattering}
To establish notation, we start by considering just elastic proton-proton scattering. In the simplest case, the high-energy elastic scattering amplitude, $T_{\rm el}$, (and correspondingly the total cross section) is parametrized by single pomeron exchange. In practice, the pre-LHC data require the trajectory of this effective 
(soft) pomeron to be
\be
\label{eq:pom-tr}
\alpha_P(t)=1+\Delta +\alpha'_P t\ ,
\ee 
with $\Delta \simeq 0.08$ and $\alpha'_P \simeq 0.25$ GeV$^{-2}$~\cite{DL}.

However, already two-particle $s$-channel unitarity generates a series of the 
non-enhanced multi-pomeron diagrams leading to the eikonal approximation.   The unitarity relaion is of the form
\be
2~{\rm Im}\ T_{\rm el}(b)= |T_{\rm el}(b)|^2+G_{\rm inel}(b)\ ,
\ee
where $G$ is the sum over all the inelastic intermediate states.
The solution gives an elastic amplitude,
\be
\label{eq:1i}
T_{\rm el}(b)~=~i(1-e^{-\Omega/2})\ ,
\ee
where one pomeron exchange describes the opacity $\Omega(s,b)$, which depends on the square of the incoming energy, $s$, and the impact parameter, $b$. Thus we have
\bea 
\sigma_{\rm tot}(s,b) & = & 2(1-{\rm e}^{-\Omega/2}), \\
\sigma_{\rm el}(s,b) & = & (1-{\rm e}^{-\Omega/2})^2, \label{eq:el}\\
\sigma_{\rm inel}(s,b) & = & 1- {\rm e}^{-\Omega}, \label{eq:inel}
 \eea
 Note, from (\ref{eq:inel}), that
 \be 
 S^2(b)\equiv e^{-\Omega}
 \label{eq:S2}
 \ee
  is the probability that  no inelastic interaction occurs at impact parameter $b$. This observation will enable us to calculate the probability that large rapidity gaps survive soft rescattering.
In terms of the opacity the elastic cross section takes the form
\be
\frac{d\sigma_{\rm el}}{dt}=\frac{1}{4\pi}  \left| \int d^2b~e^{i\q_t \cdot \b} (1-e^{-\Omega(b)/2}) \right|^2=\frac{1}{2}  \left| \int bdb~J_0(q_tb) (1-e^{-\Omega(b)/2}) \right|^2
\label{eq:el6}
\ee
where $q_t=\sqrt{|t|}$.

\begin{figure} 
\begin{center}
\vspace{-6.0cm}
\includegraphics[height=11cm]{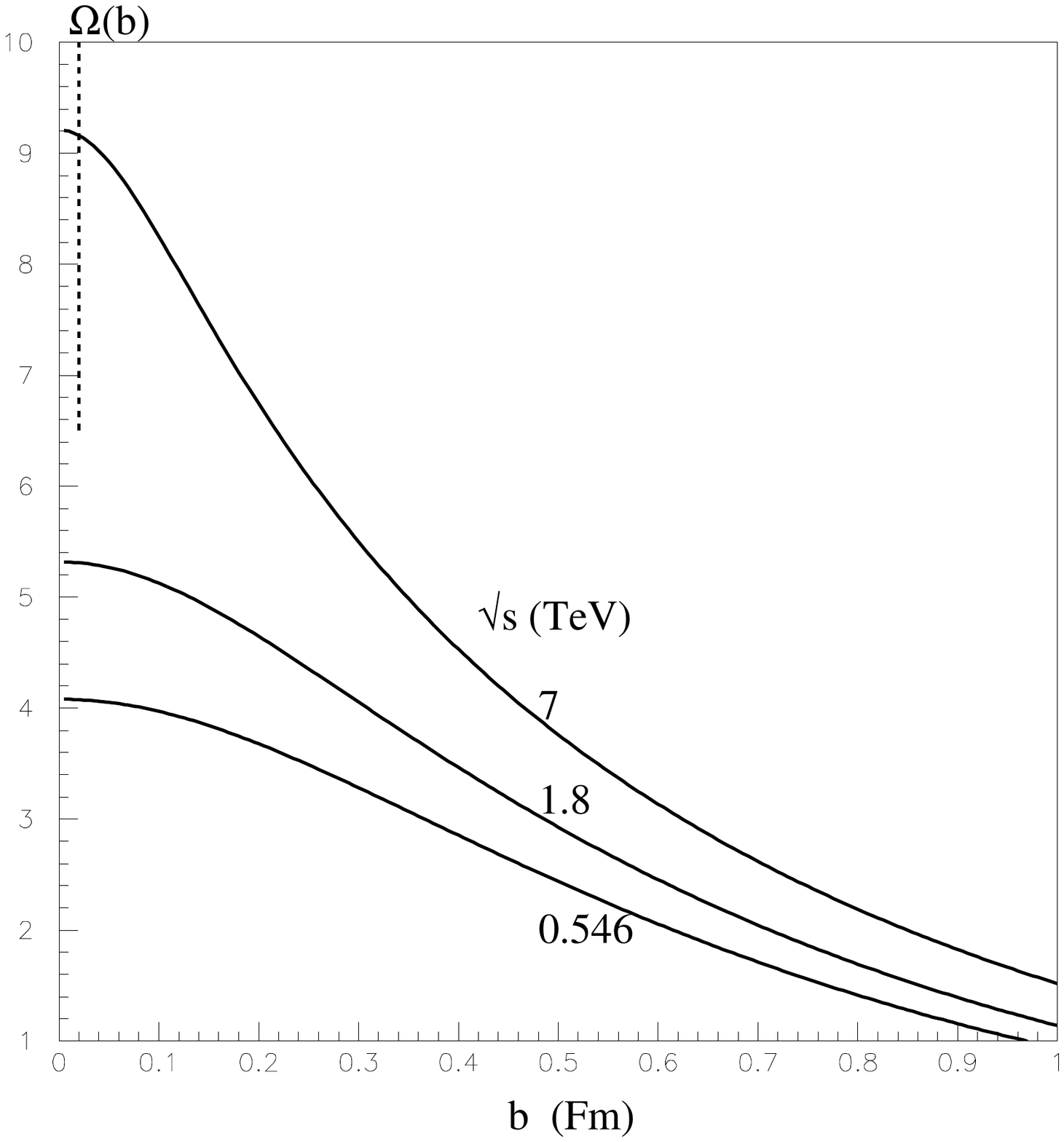}
\includegraphics[height=11cm]{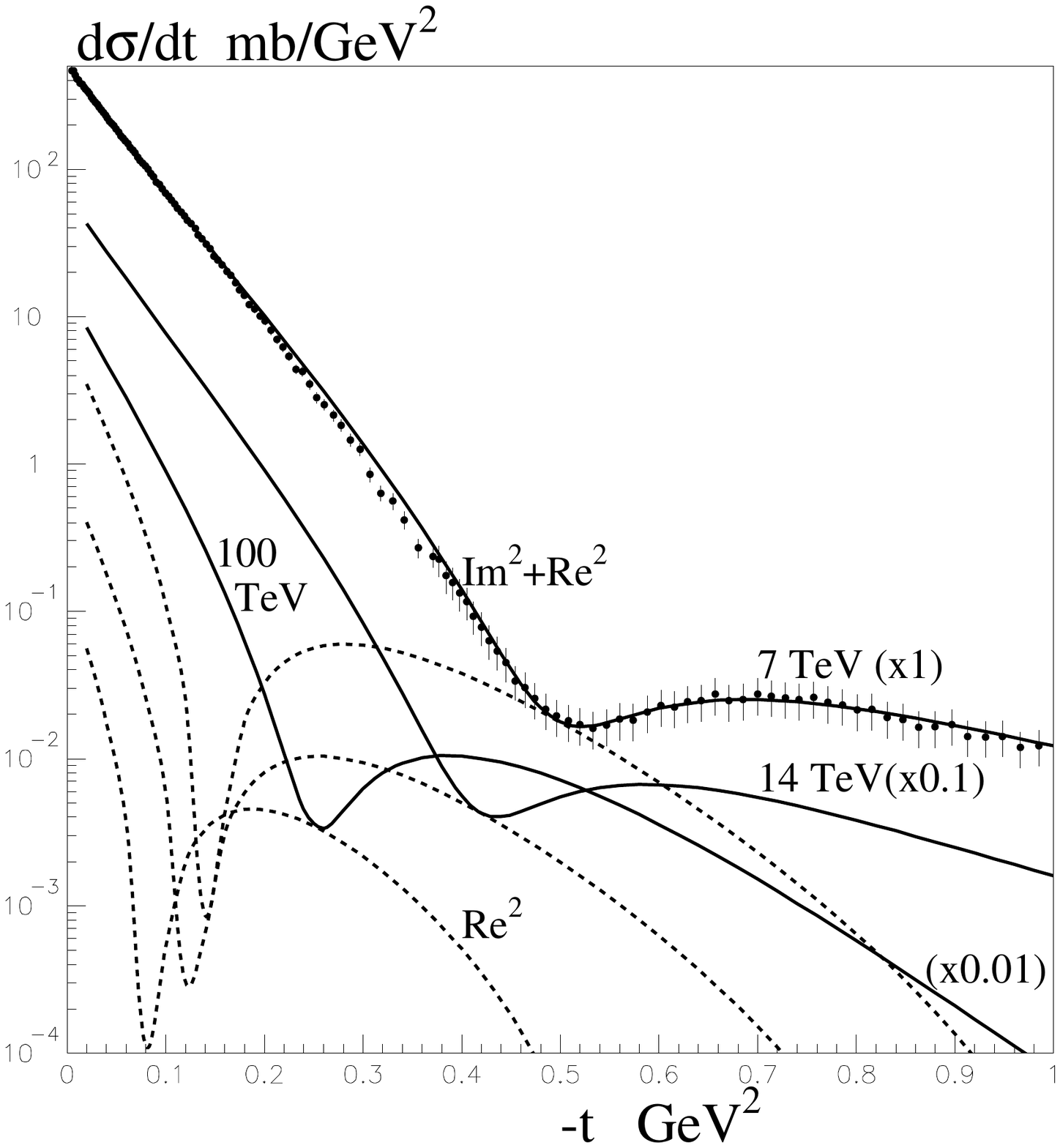}
\vspace{-0.2cm}
\caption{\sf (a) The proton opacity $\Omega(b)$ determined directly from the  $d\sigma_{\rm el}/dt$ data at 546 GeV \cite{SppS,Augier:1993sz,Arnison:1983mm}, 
1.8 TeV \cite{EEE,cdfB} and 7 TeV \cite{TOTEM2} data -- the uncertainty on the LHC value at $b=0$ is indicated by a dashed line;   (b) the continuous curves are the total contribution to the elastic cross section, and the dashed curves are the contribution of the real part of the amplitude, at three collider energies: 7, 14 and 100 TeV -- the TOTEM data at 7 TeV are shown. The figures are taken from \cite{rev}.}
\label{fig:1}
\end{center}
\end{figure}

The $b$ dependence of $T_{\rm el}$ can be obtained from the experimental data by taking the Fourier transform 
 data~\cite{Amaldi:1976gr}
\begin{equation}
{\rm Im}T_{\rm el}(b)~=~\int \sqrt{\frac{d\sigma_{\rm el}}{dt}\frac{16\pi}{1+\rho^2}}~ J_0(q_tb)~ \frac{q_tdq_t}{4\pi},
\label{eq:2i}
\end{equation}
where the square root represents Im$T_{\rm el}(q_t),$ with $\rho \equiv {\rm Re}T_{\rm el}/{\rm Im}T_{\rm el}$. In this way, we first determine Im$T_{\rm el}$ from the data for $d\sigma_{\rm el}/dt$, and then calculate $\Omega (b)$ using {(\ref{eq:1i}), assuming that $\rho$ is small. In fact, the assumption $\rho^2\ll 1$ is actually well justified, except in the diffractive dip region; see the discussion below (\ref{eq:dsel}).

 The results obtained via (\ref{eq:2i}) are shown in Fig.~\ref{fig:1}(a), where 
we compare $\Omega(b)$ obtained from elastic differential cross section data at S$p\bar{p}$S 
\cite{SppS,Augier:1993sz,Arnison:1983mm}, 
Tevatron \cite{EEE,cdfB} and LHC \cite{TOTEM2} energies. At the lower two energies the $\Omega (b)$ distributions have approximately Gaussian form, whereas at the LHC energy we observe a growth of $\Omega$ at small $b$. The growth reflects the fact that the TOTEM data indicate that we have almost total saturation at $b=0$. Note that, according to (\ref{eq:1i}), the
 value of Im$T(b=0)\to 1$ corresponds to $\Omega \to \infty$. 
Since actually we do not reach  exact saturation, the proton opacity at $b=0$ is not $\infty$, but just large numerically.  Clearly, in this region of $b$, the uncertainty on the value of $\Omega$ is large as well, see Fig.~\ref{fig:1}(a).

\section{Inclusion low-mass dissociation of the proton}
To describe proton dissociation into {\it low-mass} states we follow the Good-Walker approach \cite{GW}, and introduce so-called diffractive (or GW) eigenstates, $|\phi_i\rangle$ with $i=1,n$, that diagonalize the $T$-matrix.   The incoming `beam' and `target' proton wave functions are written as superpositions of the diffractive eigenstates
\begin{equation}
|p\rangle_{\rm beam}~=~\sum a_i |\phi_i\rangle,~~~~~~~~|p\rangle_{\rm target}~=~\sum a_k |\phi_k\rangle.
\label{eq:ai}
\end{equation}
It is sufficient to use two diffractive eigenstates, $i,k=1,2$. In terms of this $2$-channel eikonal model, we have
\be
\sigma_{\rm el}~=~  \int d^2b \left|~\sum_{i,k}|a_i|^2 |a_k|^2~(1-e^{-\Omega_{ik}(b)/2}) \right|^2,
\ee
\be
\sigma_{\rm tot}~=~  2\int d^2b~\sum_{i,k}|a_i|^2 |a_k|^2~(1-e^{-\Omega_{ik}(b)/2}) 
\label{eq:tot}
\ee
 and the `total' low-mass diffractive cross section
 \be
\sigma_{\rm el+SD+DD}~=~  \int d^2b ~\sum_{i,k}|a_i|^2 |a_k|^2 ~\left|(1-e^{-\Omega_{ik}(b)/2}) \right|^2,
\ee
where SD includes the single dissociation of both protons, DD denotes double dissociation, and where the opacity $\Omega_{ik}(b)$ corresponds to one-pomeron-exchange between states $\phi_i$ and $\phi_k$ written in the $b$-representation.
So the low-mass diffractive dissociation cross section is
\be
\sigma^{\rm D}_{{\rm low}M}~=~\sigma_{\rm el+SD+DD}-\sigma_{\rm el},
\ee
where $\sigma_{\rm el+SD+DD}$ corresponds to all
possible low-mass dissociation caused by the `dispersion' of the Good-Walker
eigenstate scattering amplitudes.

Also, the $pp$ elastic differential cross section of (\ref{eq:el6}) becomes
\be
\frac{d\sigma_{\rm el}}{dt}~=~\frac{1}{4\pi}  \left| \int d^2b~e^{i\q_t \cdot \b} \sum_{i,k}|a_i|^2 |a_k|^2~(1-e^{-\Omega_{ik}(b)/2}) \right|^2,
\label{eq:dsel}
\ee
where $-t=q_t^2$. As mentioned above, in order to correctly describe the dip region we must include the real part of the amplitude. We use a dispersion relation. For the even-signature pomeron-exchange amplitude this means
\be
A~\propto ~s^{\alpha} + (-s)^{\alpha} ~~~~~~~{\rm and~so~we ~have} ~~~~~~~\frac{{\rm Re}~A}{{\rm Im}~A}={\rm tan}(\pi(\alpha -1)/2).
\label{eq:RE}
\ee 
 Formula (\ref{eq:RE}) is transformed into $b$-space, so that the complex opacities can be constructed. For each value of $b$ we calculate $\alpha$ and determine Re$~A$ from (\ref{eq:RE}).  

The coupling $g_i$ of the pomeron to the $\phi_i$ eigenstates may be written  
\be
g_i=\gamma_i\sqrt{\sigma_0}F_i(t),
\label{eq:gi}
\ee
where the form factors satisfy $F_i(0)=1$, and $(\gamma_1+\gamma_2)/2=1$. The cross section for the interaction of eigenstates $\phi_i$ and $\phi_k$, via one-pomeron-exchange, is
$\sigma_{ik}=\sigma_0 \gamma_i \gamma_k (s/s_0)^{\alpha_P-1}$.
The parameters of the form factors, $F_i(t)$, are adjusted to give, among other things, the observed behaviour of $d\sigma_{\rm el}/dt$
versus $t$, as shown in Fig. \ref{fig:1}(b).

 \section{High-mass dissociation   \label{sec:hm}   }
 In the absence of absorptive corrections, the  $pp\to  p+X$ cross section for dissociation into a system $X$ of high mass $M$ is
\be
\frac{M^2 d\sigma_{\rm SD}}{dtdM^2}~=~\frac{g_{3P}(t)g_N(0)g_N^2(t)}{16\pi^2}~\left(\frac{s}{M^2}\right)^{2\alpha(t)-2}~\left(\frac{M^2}{s_0}\right)^{\alpha(0)-1},
\label{eq:3P}
\ee
where $g_N(t)$ is the coupling of the pomeron to the proton and $g_{3P}(t)$ is the triple-pomeron coupling.
The value of the coupling $g_{3P}$ is obtained from a triple-Regge analysis of lower energy data. However, $g_{3P}$ is actually an effective vertex with coupling
\be
g_{\rm eff}~=~g_{3P}~\langle S^2\rangle
\ee
which already includes the suppression or survival factor $S^2(b)=\exp(-\Omega(b))$ -- the probability that no other secondaries, simultaneously produced in the same $pp$ interaction, populate the rapidity gap region, see (\ref{eq:S2}).  A more precise analysis \cite{Luna} accounts for the survival effect $S^2_{\rm eik}$ caused by the eikonal rescattering of the fast `beam' and `target' partons.  In this way, a coupling $g_{3P}$ about a factor of 3 larger than $g_{\rm eff}$ is obtained, namely 
\be
g_{3P} \equiv \lambda g_N~~~~~{\rm with}~~~~~\lambda\simeq 0.2,
\label{eq:lamb}
\ee
where $g_N$ is the coupling of the pomeron to the proton.

\section{Experimental evidence of the $k_t(s)$ effect   \label{sec:kts}}
Can a simultaneous `global' description be obtained, of all the high-energy elastic and diffractive data, accounting for all the puzzles listed in Section \ref{sec:items}?  This was the objective of \cite{kmr2}, which was based on earlier observations in  \cite{KMR-s3,kmr1}.  This analysis was reviewed in detail in \cite{rev}.
Here we sketch the main results.

Loosely speaking, one of the main puzzles, in the data sets, is that the diffractive dissociation cross sections observed at the LHC are smaller than expected from conventional RFT extrapolations of lower energy collider data. There is a good physics reason why this is so -- the $k_t(s)$ effect \cite{kmr1,kmr2,rev}.

\subsection{Energy dependence of low-mass dissociation}
We discuss, first, item (b) on the list; that is, the energy dependence of 
 $\sigma_{\rm D}^{{\rm low}M_X}$. The pomeron-$|\phi_i \rangle$ coupling, $g_i$, 
 is driven by the impact parameter separation, $\langle r\rangle$, between the partons in the $|\phi_i\rangle$ state. The well known example is so-called colour transparency, where the cross section $\sigma\propto \alpha_s^2\langle r^2\rangle$. However, if the transverse size of the pomeron, ($\sim 1/k_t$),  becomes much smaller than this separation, then the cross section (and coupling) will be controlled by the pomeron size, that is by the characteristic $k_t$ in the pomeron ladder, which we denote by $k_P$. In this limit $\sigma\propto 1/k^2_P$.  Therefore it is natural to choose the following parametrization for the $\gamma_i$ in the pomeron-$|\phi_i\rangle$ couplings of (\ref{eq:gi})
\be
\gamma_i\propto \frac 1{k^2_P+k^2_i}, 
\label{eq:gam-i}
\ee 
where $k_P^2$ is expected to have an energy dependence
\be 
k^2_P (s)= k^2_{P0}\left(\frac{s x^2_0}{s_0}\right)^D\ .
\label{eq:D}
\ee 
In other words, during the evolution in $\ln(1/x)$, due to the 
diffusion in $\ln k^2_t$,  $k^2_P$ grows approximately as a power $D$ of $1/x$. Of course, we do not expect that the whole available $\ln(1/x)$ (rapidity) space will be subject to  diffusion. Rather, we assume, that as $x$ decreases, the diffusion starts from some relatively low $x=x_0$ parton with $x_0=0.1$. That is, the rapidity space available for the $\ln k^2_t$ diffusion is not  $\ln(s/s_0)$, but is diminished by $\ln(1/x_0)$ from both sides. (As usual we use  $s_0=1$ GeV$^2$.) The typical transverse momentum of this (starting) parton, inside the state $\phi_i$, is denoted by the parameter $k_i$ in (\ref{eq:gam-i}). 

The parametrisation of $\gamma_i$ in (\ref{eq:gam-i}) is such that at very large energies all the $\gamma_i\to 1$. That is, the interaction will not destroy the coherence of the wave functions of the incoming protons. In other words, $\sigma_{\rm D}^{{\rm low}M_X}$ will decrease with energy, while at lower energies we tend to the naive expectation $\gamma_i \propto 1/k^2_i$. In the `global' description \cite{kmr2}, the dissociation is slowed sufficiently with increasing energy such that the values of the cross section $\sigma_{\rm D}^{{\rm low}M_X}$ are compatible with the data -- namely, the model gives 2.6 mb at $\sqrt{s}=62.5$ GeV, and 3.8 mb at $\sqrt{s}=7$ TeV.  The latter value is consistent with the TOTEM measurement of 2.6$\pm$2.2 mb.

\subsection{Energy dependence of high-mass dissociation}
An analogous puzzle was listed in item (c) of Section \ref{sec:items}, so it is natural to introduce some energy dependence into the triple pomeron coupling $\lambda$, of (\ref{eq:lamb}), which drives high-mass dissociation.  As before, it arises from
the characteristic transverse momenta of the intermediate partons inside the pomeron ladder (i.e. the size of the pomeron), which depend on the rapidity (or energy) of corresponding partons, It looks natural to take 
\be
\lambda\propto 1/k^2_P(s)\ ,
\label{eq:lambda1}
\ee
where $k_P^2=\langle k_t^2 \rangle$. The diffusion in ln$k_t^2$ occurs from both the beam and target sides of the ladder. Following (\ref{eq:D}), we take $k_P^2 \propto (x_0/x)^D$ for diffusion from one side and $k_P^2 \propto (x_0/x')^D$ from the other side; where to evaluate $x'$ we use the relation $xx's=\langle m^2_T\rangle$, and assume $\langle m^2_T\rangle = s_0=1~\GeV^2$.
So we parametrize $k_P(s)$ by
\be
k^2_P (s)=k^2_0
 \left(\left(\frac{x_0}x\right)^D+\left(\frac{x_0}{x'}\right)^D\right)\ ,
\label{eq:lambda2}
\ee 
where we take the same parameter $D$ and evolve from the same starting point $x_0=0.1$ as (\ref{eq:D}) for $\gamma_i$ of (\ref{eq:gam-i}). We calculate  $x'$ as $x'=s_0/xs$ with $s_0=1$ GeV$^2$.  If $x>x_0$ then $x_0/x$ ratio is replaced by 1, and similarly for $x'$.  The global description \cite{kmr2} gives $D=0.28$.

With this energy dependence of the coupling $\lambda$, the values obtained in \cite{kmr2,rev} for the single proton 
 dissociation cross section (integrated over the three mass intervals used by TOTEM \cite{TO4}) are shown in Table \ref{tab:1}. 
\begin{table} [h]
\begin{center}
\begin{tabular}{|l|c|c|c|}\hline
 Mass interval (GeV) &   (3.4,~8) & (8,~350)&  (350, 1100)   \\ \hline
  Prelim. TOTEM data & 1.8 & 3.3  & 1.4   \\
 CMS data &  &  4.3 &  \\
 Model \cite{kmr2} & 2.3  & 4.0 & 1.4 \\
 \hline

\end{tabular}
\end{center}
\caption{\sf The values of the cross section (in mb) for single proton dissociation (integrated over the three    indicated mass intervals) as observed by TOTEM \cite{TO4}, compared with the values obtained in the model of \cite{kmr2,rev}. Recall that TOTEM claims that their preliminary measured cross sections have a 20\% uncertainty. Note that the value quoted for the CMS \cite{CMSdiff} cross section of dissociation is integrated over the 12 - 394 GeV $M_X$ interval (close to, but in terms of $\ln M_X$, a bit smaller than, the interval (8 - 350) GeV chosen by TOTEM).}
\label{tab:1}
\end{table}
We see that the agreement with the mass dependence of the TOTEM data is now satisfactory. 
Moreover, note that $\lambda=0.18$ at relatively low energies, when both $x>x_0$ and $x'>x_0$, such that $\lambda$ ceases to be energy dependent. This is in agreement with the value 0.2 of (\ref{eq:lamb}) obtained in the triple-Regge analysis.

\subsection{An interesting and unusual effect on the energy behaviour of $\sigma_{\rm tot}$}
Allowing the pomeron couplings to be energy dependent produces the unusual energy behaviour of the total cross section noted in item (a) of Section \ref{sec:items}. Indeed, the observation that the $\gamma_i\to 1$ at high energy, means that the growth of the total cross section speeds up in the CERN-ISR $\to$ LHC $\to$ 100 TeV interval.  For example, if we followed the conventional `classic' Regge approach and assumed that the couplings $\gamma_i$ were independent of energy, and fix the values corresponding to the CERN-ISR energy of $\sqrt s=50$ GeV, then we obtain $\sigma_{\rm tot}=74,\ 91$ and 134 mb for $\sqrt s=1.8,\ 7$ and 100 TeV respectively.  On the other hand, allowing the couplings to be energy dependent, as follows from (\ref{eq:D}), the  values of $\sigma_{\rm tot}$ grow faster with energy, namely. 77, 99 and 166 mb, and now in accord with the LHC measurement.

These numbers demonstrate an important new fact. That is, that the energy dependence of the total and elastic cross sections is not only driven by the parameters of the pomeron trajectory, but also by the energy behaviour of the factors $\gamma_i$; in other words by the decomposition of the proton-pomeron coupling between the different Good-Walker eigenstates. Note that this `acceleration' of the total cross section growth, due to the variation $\gamma_i \to 1$, takes place in only one energy interval. We are fortunate to observe it in just the  S$p\bar{p}$S -- LHC collider interval.

\subsection{ Factorisation for the diffractive cross sections?}
Concerning item (d), we note that TOTEM have measured $\sigma_{\rm el},~\sigma_{\rm SD}$ and $\sigma_{\rm DD}$ in some kinematic interval. If we take the simplest pomeron exchange diagrams that describe these processes, then we obtain the naive factorisation relation~
$\sigma_{\rm DD}\sigma_{\rm el}/(\sigma_{\rm SD})^2 = 1$, whereas the data give ~$0.116\times 25/(0.9)^2 \simeq 3.6$, integrated over the same kinematic region.
Here $\sigma_{\rm SD}$ is the single dissociation cross section from {\it one} proton, not the sum of both dissociations.  This discrepancy is not a surprise.  We expect violation of the naive result due to the different $t$-slopes and survival factors for the processes.  These, together with a more careful treatment of the calculation of $\sigma_{\rm DD}$, removes the discrepancy \cite{kmr2}.

\section{Tension between the high-mass single dissociation data}
Although TOTEM have made the most detailed observations of high-mass single proton dissociation in high energy $pp$ collisions, the  `global' diffractive  model \cite{kmr2} was tuned to simultaneously describe the TOTEM data {\it together} with earlier measurements of single dissociation. 
\begin{figure} 
\begin{center}
\vspace{-6.cm}
\includegraphics[height=11cm]{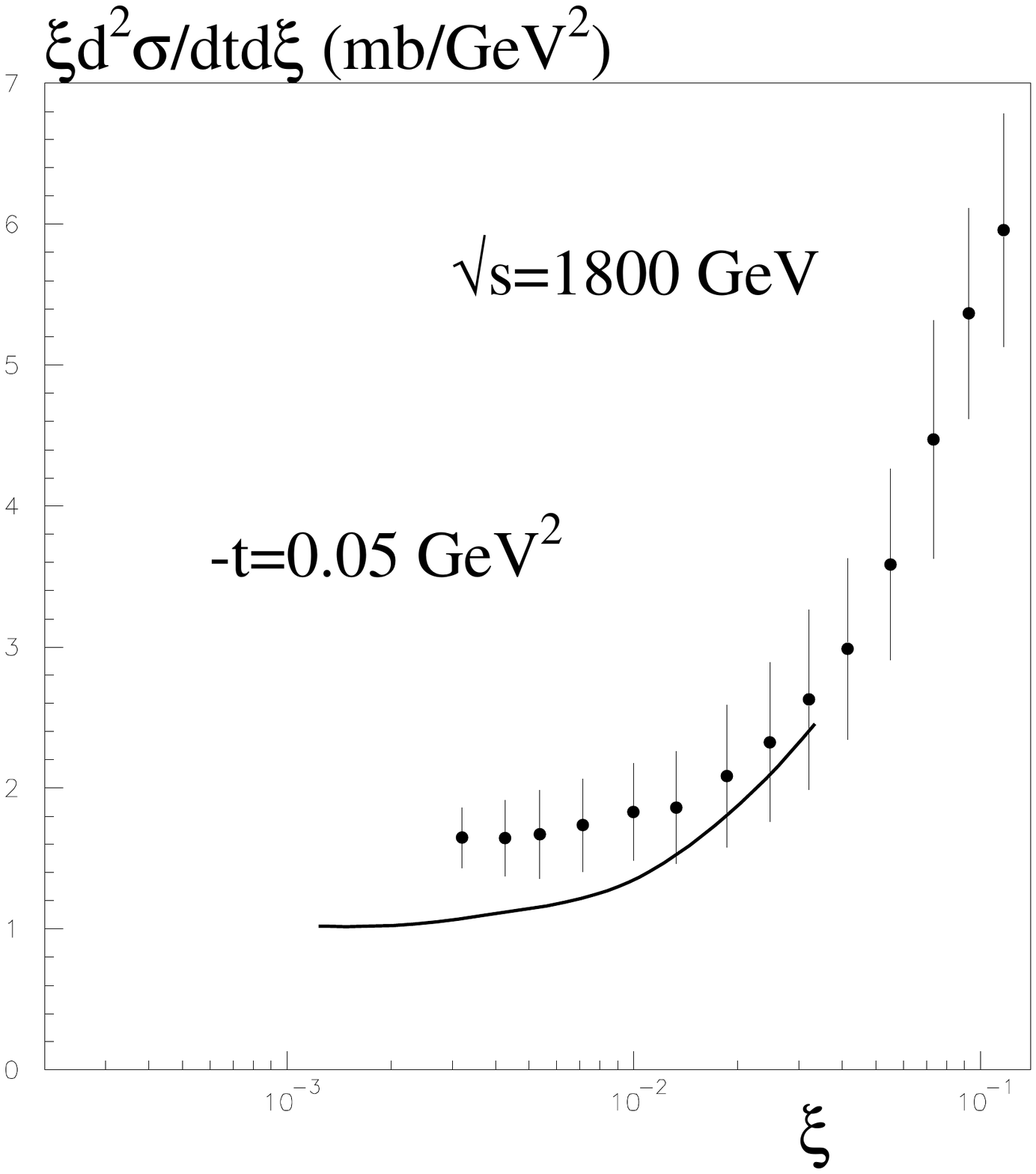}
\includegraphics[height=11cm]{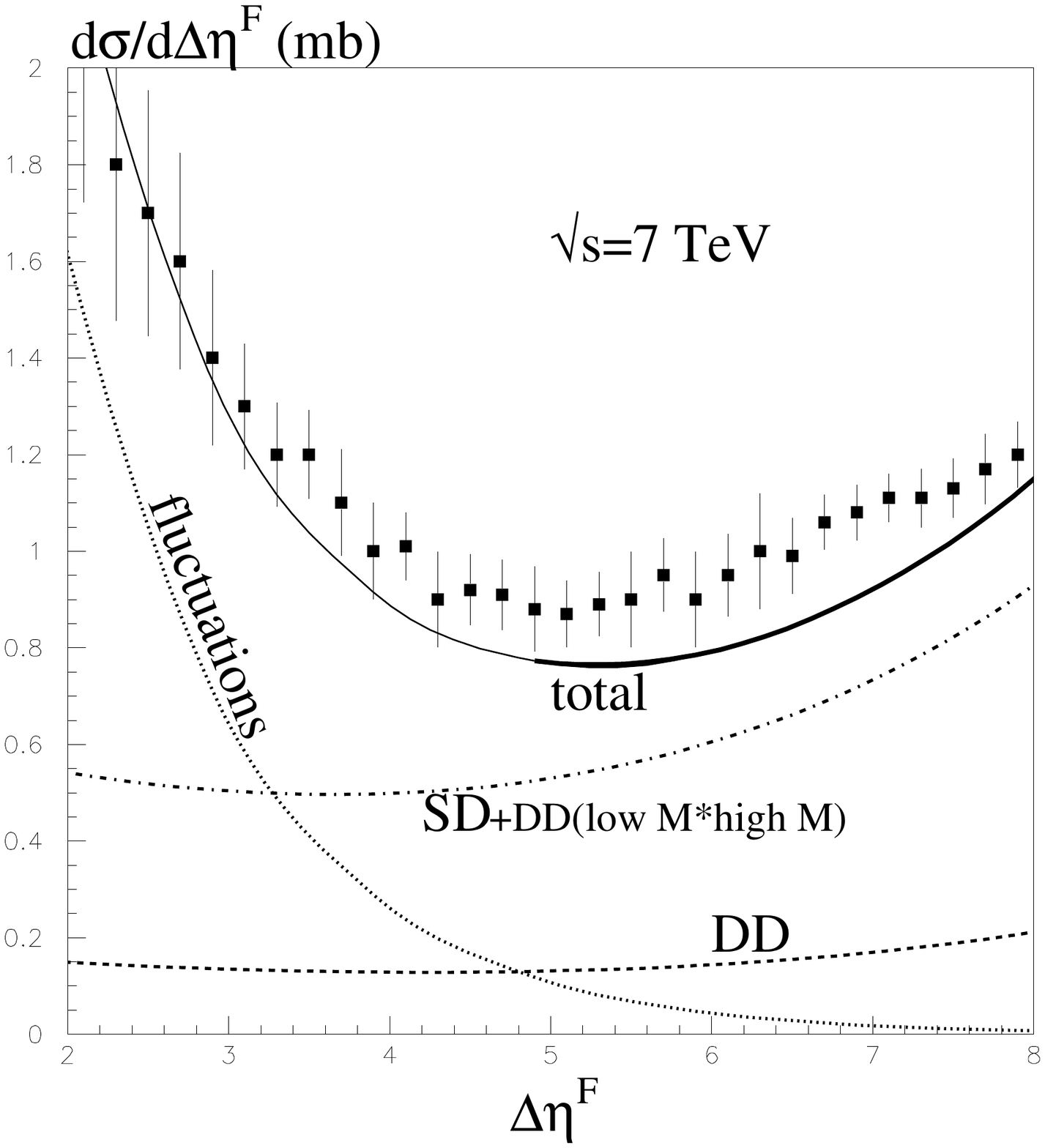}
\vspace*{-.5cm}
\caption{\sf (a) The comparison of the model \cite{kmr2} with data for single proton dissociation measured by the CDF collaboration, given in \cite{GM} but not including a normalisation uncertainty of about 10-15\%. The inclusion of the secondary Reggeon contribution RRP is responsible for the rise of the curve as $\xi$ increases. (b) The ATLAS \cite{atl} measurements of the inelastic cross section differential in rapidity gap size $\Delta\eta^F$ for particles with $p_T>200$ MeV. Events with small gap size ($\Delta\eta^F \lapproxeq 5$) may have a non-diffractive component which arises from fluctuations in the hadronization process \cite{FZ}. This component increases as $\Delta\eta^F$ decreases (or if a larger $p_T$ cut is used \cite{FZ,atl}).  
The data with $\Delta\eta^F\gapproxeq 5$ are dominantly of diffractive origin, and are compared with the `global' diffractive model \cite{kmr2}. The DD contribution of events where both protons dissociate, but the secondaries produced by one proton go into the beam pipe and are not seen in the calorimeter, is shown by the dashed curve. The figures are taken from \cite{kmr2}.}
\label{fig:2}
\end{center}
\end{figure}
Formally the diffractive dissociation data from different groups do not contradict each other, since they are measured for different conditions. However, global fits of all diffractive data reveal the tensions between the data sets \cite{kmr2,rev,Ost1}. 
 Indeed, it is not easy to reconcile the preliminary TOTEM measurements for $\sigma_{\rm SD}$ with (a) the `CDF' measurement of single dissociation and (b) the yield of Large Rapidity Gap events observed by  ATLAS and CMS.  Basically the global description of \cite{kmr2} overestimates the TOTEM data and underestimates the CDF \cite{GM}, ATLAS \cite{atl} and CMS \cite{CMSdiff} data. This can be seen in Fig. \ref{fig:2} and from Table 1.

\section{Discussion}
We have discussed a model \cite{kmr2} in which all the high-energy  diffractive data may simultaneously be described within the Regge Field Theoretic framework based on only one pomeron pole. However, to reach  agreement with the data, the model includes pomeron-pomeron interactions, arising from multi-pomeron vertices, and allows for the $k_t(s)$ dependence of the multi-pomeron vertices. Recall that, due to the BFKL-type diffusion in $\ln k^2_t$ space,  together with the stronger absorption of low $k_t$ partons, the typical transverse momentum, $k_t$, increases with energy depending on the rapidity
 position of the intermediate parton or the multi-pomeron vertex. This $k_t(s)$ effect enables the model to achieve a relatively low probability of low-mass 
 dissociation of an incoming proton and to reduce the cross section of high-mass  
 dissociation in the central rapidity region in comparison with that observed closer 
 to the edge of available rapidity space -- both of which are features demanded by the recent TOTEM data.

The model \cite{kmr2} can be refined when more precise and consistent diffractive data become available. At present the parameters are {\it tuned}
 to give a reasonable description of all aspects of the available diffractive data.  If, instead, a $\chi^2$-fit to the  data had been performed, then the few dissociation measurements of TOTEM (the preliminary values of $\sigma_{\rm SD}$ with 20\% errors in three mass intervals, and one value of $\sigma_{\rm DD}$) would have carried little weight.
On the  other hand, all the TOTEM data are self-consistent between themselves. Moreover, these data  reveal a very reasonable tendency of the $d\sigma_{\rm SD}/d\xi$ dependence, close to that predicted in \cite{KMR-s3} where the $k_t$ distribution of the intermediate partons inside the pomeron ladder, and the role of the transverse size of the different QCD pomeron components, were accounted for  more precisely.
 Therefore, the model \cite{kmr2} was tuned to achieve a `compromised' description of the data sets in `tension'.  On the theory side, the model could be improved by including
a more complicated (non-local) structure of the original pomeron. At present, the model considers just an `effective' pomeron pole renormalized by enhanced absorptive corrections.\footnote{The model (which includes eikonalized absorption, but not enhanced absorption)  has an effective pole with a trajectory $\alpha_P(t) \simeq 1.12+0.05t$, intermediate between the effective soft pomeron with  $\alpha_P(t)\simeq 1.08+0.25t$ of (\ref{eq:pom-tr}) and the `bare' QCD pomeron with $\alpha_P(t)\simeq 1.25+0t$, in GeV units.}

\section*{Acknowledgements}
I thank Olga Piskunova for arranging a very fruitful and enjoyable Workshop, and Misha Ryskin and Valery Khoze for many enjoyable collaborations and valuable discussions on the subject of the article.

\section*{References}

\bibliographystyle{iopartnum}
\bibliography{F}

\end{document}